\documentclass[twocolumn]{aastex631}
\usepackage[utf8]{inputenc}
\usepackage[T1]{fontenc}
\usepackage{graphicx} 
\usepackage{amsmath}
\usepackage[sort&compress]{natbib}
\usepackage{float}
\usepackage{xspace}

\newcommand{\degs}[1]{#1^{\circ}}

\newcommand{\gray}{$\gamma$-ray\;}
\newcommand{\sv}{\langle \sigma v \rangle}
\newcommand{\fermi}{\textit{Fermi}\xspace}

\usepackage{hyperref}
\usepackage{tablefootnote}

\begin{document}
\title{Evaluating the Potential to Constrain Dark Matter Annihilation with \textit{Fermi}-LAT Observations of Ultra-Faint Compact Stellar Systems}

\author{Antonio Circiello}
\affiliation{Department of Physics and Astronomy, 
Clemson University, 4
Clemson, SC 29631, USA}
\author{Alex McDaniel}
\affiliation{Department of Physics and Astronomy, 
Clemson University, 
Clemson, SC 29631, USA}
\author[0000-0001-8251-933X]{Alex Drlica-Wagner}
\affiliation{Fermi National Accelerator Laboratory, P.O.\ Box 500, Batavia, IL 60510, USA}
\affiliation{Kavli Institute for Cosmological Physics, University of Chicago, Chicago, IL 60637, USA}
\affiliation{Department of Astronomy and Astrophysics, University of Chicago, Chicago, IL 60637, USA}
\author{Christopher Karwin}
\affiliation{NASA Postdoctoral Program Fellow, 
NASA Goddard Space Flight Center, 
Greenbelt, MD, 20771, USA}
\author[0000-0002-6584-1703]{Marco Ajello}
\affiliation{Department of Physics and Astronomy, 
Clemson University, 
Clemson, SC 29631, USA}
\author[0000-0003-2759-5625]{Mattia Di Mauro}
\affiliation{Istituto Nazionale di Fisica Nucleare, Sezione di Torino, 
Via P. Giuria 1, 10125 Torino, Italy}
\author[0000-0002-3849-9164]{Miguel Á. Sánchez-Conde}
\affiliation{Instituto de F\'isica Te\'orica UAM-CSIC, Universidad Aut\'onoma de Madrid, C/ Nicol\'as Cabrera, 13-15, 28049 Madrid, Spain}
\affiliation{Departamento de F\'isica Te\'orica, M-15, Universidad Aut\'onoma de Madrid, E-28049 Madrid, Spain}

\begin{abstract}
    Recent results from numerical simulations and models of galaxy formation suggest that recently discovered ultra-faint compact stellar systems (UFCSs) in the halo of the Milky Way (MW) may be some of the smallest and faintest galaxies.
    If this is the case, these systems would be attractive targets for indirect searches of weakly interacting massive particle (WIMP) dark matter (DM) annihilation due to their relative proximity and high expected DM content.
    In this study, we analyze 14.3 years of gamma-ray data collected by the \textit{Fermi}-LAT coincident with 26 UFCSs. 
    No significant excess gamma-ray emission is detected, and we present gamma-ray flux upper limits for these systems. 
    Assuming that the UFCSs are dark-matter-dominated galaxies consistent with being among the faintest and least massive MW dwarf spheroidal (dSphs) satellite galaxies,  we derive the projected sensitivity for a dark matter annihilation signal. 
    We find that observations of UFCSs have the potential to yield some of the most powerful constraints on DM annihilation, with sensitivity comparable to observations of known dSphs and the Galactic center.
    This result emphasizes the importance of precise kinematic studies of UFCSs to empirically determine their DM content. 
    \end{abstract}

\section{Introduction}\label{sec:intro}
The Milky Way (MW) is surrounded by more than fifty dwarf spheroidal (dSph) satellite galaxies that reside in dark matter (DM) halos \cite[e.g.,][ and references therein]{Simon+2019}. 
While it was once possible to distinguish DM-dominated galaxies from DM-deficient star clusters based on size and luminosity \cite[e.g.,][]{Willman:2005}, many recently discovered systems have sizes and luminosities that blur that boundary \citep[see][and references therein]{Cerny+2023b}.
In particular, dozens of ultra-faint compact stellar systems (UFCSs) have been discovered at distances of several tens of kpc possessing low luminosities, $L_\star < 10^4 L_\odot$, and small physical sizes, $r_h < 30$\,pc (see Table~\ref{tab:sample} for a list of references).
Detailed kinematic studies of these systems are challenging due to their distances and low luminosities; however, recent theoretical arguments suggest that these systems could be the faintest and least massive DM-dominated galaxies \cite[e.g.,][]{Manwadkar+2022,Errani2023a}.  
One example in this class of systems is the recently discovered satellite Ursa Major III/UNIONS I \citep[UMa\;III,][]{smith2023}.
While measurements of the velocity dispersion of this system are inconclusive due to the small number of measured stars, simulations suggest that a DM halo is necessary to stabilize the system against tidal dissolution \citep{Errani2023}. 
If this system is indeed a DM-dominated galaxy, then galaxy--halo modeling arguments suggest that its host DM halo had a peak mass of $\lesssim 10^8 M_\odot$ \citep{Sawala+2015, Nadler:2020, Manwadkar+2022, Revaz:2023}. 
This work is based on the assumption that UMa III and other UFCSs (see Sec.~\ref{sec:sample}) are DM-dominated and that their subhaloes are similar in nature to the dSphs \citep{Manwadkar+2022}.
This second hypothesis is needed since the size of a subhalo can vary strongly depending on its tidal history. 
\cite{Errani2023a} show that the subhaloes hosting the UFCSs could have similar extensions to the dSphs, or they could have been stripped by tides down to their stellar components.
In order to evaluate the projected sensitivity to DM annihilation signals, we start by assuming all objects in our sample are similar in nature to the dSphs.
We then evaluate how the sensitivity is affected if a subset of the targets contain much less DM than is implied by our initial assumption (see Sec. \ref{sec:ResDisc}).

Under these hypotheses, the UFCSs would be powerful targets for searches of \gray emission from the annihilation of weakly interacting massive particles \citep[WIMPs, for reviews, see][]{Jungman+1996, Bergstrom+2000, Abdo+2010_2} DM.
UFCSs are an attractive target primarily due to their proximity, which could lead to a higher \gray flux from DM annihilation compared to the MW satellites studied previously.
The detection of excess \gray emission coincident with one or more of these systems (especially the closest ones) would be exciting even if their DM contents have not yet been measured.
On the other hand, a null detection of nearby DM-dominated UFCSs would increase the sensitivity of existing studies of dSphs, which already yield the most powerful and robust constraints on DM annihilation \citep[][and references therein]{mcdaniel2023}.

In this Letter, we present the results of a search for \gray emission coincident with a sample of 26 UFCSs.
Our sample was selected to be consistent with the galaxy size--luminosity relationship derived in \citet{Manwadkar+2022}. 
Our analysis of the \textit{Fermi}-LAT \gray data adopts the procedure developed for the analysis of dSphs by  \citet{mcdaniel2023}, which closely follows previous LAT analyses of dSphs \citep[e.g., ][]{Ackermann+2015,Albert+2017}.
We find no significant excesses of \gray emission coincident with any of the UFCSs, and we present flux upper limits for each system.
Under the assumption that the UFCSs are DM-dominated systems consistent with the known population of dSphs, we proceed to make projections for the ability to constrain the DM annihilation cross-section from the null detection of \gray emission.
We also derive sensitivity projections for a subset of 17 out of the 26 UFCSs from the initial sample, applying a more stringent selection in size--luminosity space.
We find that, although subject to large uncertainties at present, the DM cross-section upper limits derived under these assumptions for both samples of UFCSs are more sensitive than those obtained from recent dSphs analyses \citep[e.g.,][]{DiMauro+23, mcdaniel2023}, emphasizing the importance of kinematic studies of UFCSs to empirically determine their DM content. 

The Letter is organized as follows.
In Sec.~\ref{sec:sample} we discuss how our samples of UFCSs were selected.
In Sec.~\ref{sec:DM} we detail the assumed model for the DM halos of UFCSs as well as the DM annihilation channels considered.
Sec.~\ref{sec:analysis} is devoted to the \fermi-LAT data selection and the analysis procedure.
Finally, in Sec.~\ref{sec:ResDisc}, we discuss the results and conclude.

\section{Sample selection}\label{sec:sample}
\begin{figure}
    \centering
    \includegraphics[width = \linewidth]{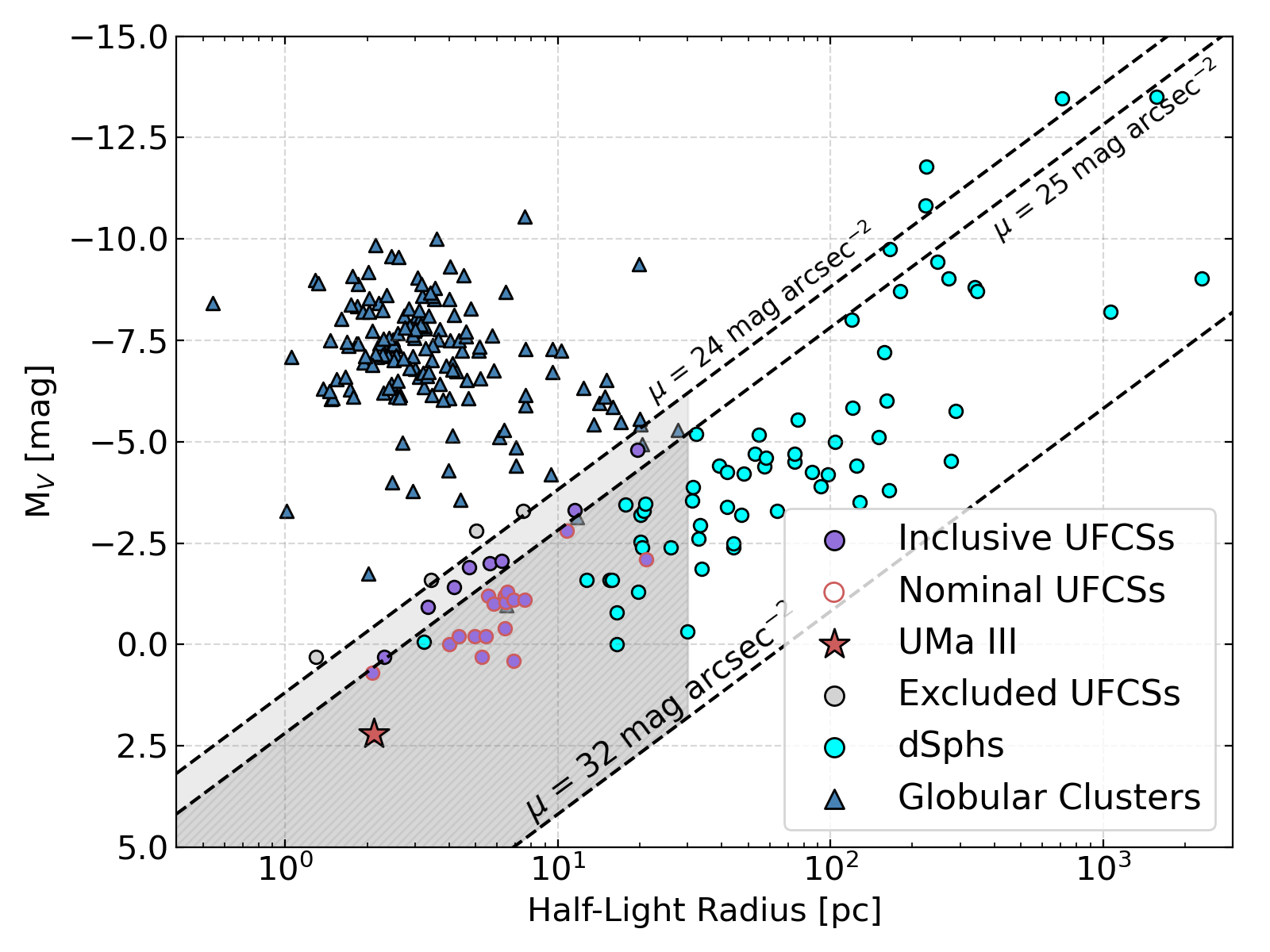}
    \caption{$V$-band absolute magnitude vs.\ physical half-light radius for Milky Way globular clusters, dSphs, and UFCSs. 
    The dashed lines mark the minimum surface brightness used to select the inclusive ($\mu > 24 \mathrm{\, mag \,arcsec}^{-2}$) and nominal ($\mu > 25 \mathrm{\, mag \,arcsec}^{-2}$) samples, based on the model of \citet{Manwadkar+2022}. The light and dark grey shaded areas represent the regions of inclusive and nominal selections, respectively. The purple points are the UFCSs in the inclusive sample. The UFCSs that are also included in the nominal sample are circled in red. A complete list of both samples is reported in Table~\ref{tab:sample}. The grey points are the UFCSs that fall outside our selections and were excluded from the analysis for this reason. The cyan points are the confirmed dSphs from \texttt{dmsky} (\url{https://github.com/fermiPy/dmsky}). The dark blue triangles are the confirmed globular clusters from the Local Volume Database \citep{Pace:2024}.}
    \label{fig:mv_hlr}
\end{figure}

The populations of DM-dominated dSphs and DM-deficient classical globular clusters have historically been separable in the space of size (as indicated by the half-light radius) and luminosity (Fig.~\ref{fig:mv_hlr}).
However, as the sensitivity of optical imaging surveys has increased, fainter and more compact systems have been discovered populating a region of parameter space where their classification is uncertain \citep[e.g.,][]{Simon+2019, Drlica-Wagner:2020}.
Furthermore, the low luminosities and distances of these systems make it difficult to measure the velocities of enough member stars to confidently determine velocity dispersions and dynamical masses.
Early observational studies assumed that these systems were a low-luminosity tail of the globular cluster population, possibly formed and accreted within satellite galaxies that were subsequently disrupted \citep[e.g.,][]{Mau+2019}.
However, recent theoretical modeling work has suggested that the population of dwarf galaxies may extend to equivalently small sizes \citep[e.g.][]{Manwadkar+2022,Errani2023a}.
In particular, the regulator model of galaxy formation developed in \citet{Kravtsov:2022} and applied to simulations of a Milky-Way-like system in \citet{Manwadkar+2022} predicts that a significant population of compact, low-luminosity satellite galaxies exists.

We select a sample of UFCSs that are consistent in physical size and absolute magnitude with the locus of galaxies produced by the model of \citet[][see their Fig.~12]{Manwadkar+2022}.
In particular, we select UFCSs that have azimuthally averaged physical half-light radii $r_{1/2} < 30$\,pc and surface brightnesses of $24 \mathrm{\, mag \,arcsec}^{-2} < \mu_V < 32 \mathrm{\,  mag \, arcsec}^{-2}$. 
Furthermore, we remove systems that have previously been included in population studies of \gray emission from dSphs (Bootes V, Cetus II, Draco II, Grus I, Leo Minor I, Pictor I, Segue 1, Segue 3, Triangulum II, Tucana V, Virgo II, and Willman 1) \citep[e.g.,][]{Albert+2017, mcdaniel2023}, or have been confirmed to be DM-deficient from kinematic measurements \citep[AM 4, Palomar 5, Palomar 13, Palomar 14;][2010 edition]{Harris+1996}.
We apply the same selection criteria used for the dSphs in \citet{mcdaniel2023} to avoid contamination from other sources of \gray emission by checking if any targets in our sample fall within the 95\% confidence radius of a 4FGL-DR4 source, or within $0.1\degs{}$ \citep[i.e., roughly the mean of the 95\% confidence radius of point sources in the 4FGL,][]{4fgl-dr4} of a source in the BZCat, CRATES, and WIBRaLS catalogs. No targets were removed by this criteria.
We refer to UFCSs passing this selection as the `inclusive sample', which contains 26 targets (Table \ref{tab:sample}).
We also analyze a subset of this sample, selecting targets in a more restrictive range of surface brightness ($25 \mathrm{\, mag \,arcsec}^{-2} < \mu_V < 32 \mathrm{\,  mag \, arcsec}^{-2}$), which corresponds to the region of maximum density of galaxies predicted by \cite{Manwadkar+2022}. 
We refer to this selection as our `nominal sample', which contains 17 of the 26 UFCSs from the inclusive sample. 
Targets that fall into this selection are listed separately in Table~\ref{tab:sample}.

\section{Dark Matter Annihilation}\label{sec:DM}

Astrophysical searches for DM involve looking for the signatures of DM interactions.
In the WIMP DM model, one possible signature of DM annihilation is \gray emission through the production of high-energy standard model particles \citep[e.g., ][for a review]{Bertone+2005}.
The expected \gray flux from DM annihilation is \citep{Bergstrom+1998}:
\begin{equation}
\label{eq:DM_flux}
    \frac{d\Phi_{\chi}}{dE} = J \times \frac{1}{4\pi}\frac{\sv}{2 M_\chi^2} \sum_i \beta_i \frac{dN_i}{dE},
\end{equation}
where $M_\chi$ is the rest mass of the DM particle, and $\sv$ is the velocity-averaged DM annihilation cross-section. 
The sum is performed over the annihilation channels, where $\beta_i$ is the branching ratio for the \textit{i}-th channel.
The \gray spectrum per annihilation event given the annihilation channel, $dN/dE$, is given by the DM model that is being considered.
In this work, we use the PPPC4DMID\footnote{http://www.marcocirelli.net/PPPC4DMID.html} tables from \citet{Cirelli2011} to compute the different annihilation spectra.
The annihilation channels considered are $b\Bar{b}$ and $\tau^{+}\tau^{-}$.\footnote{In a recent work by \cite{Arina:2023eic} the authors have computed updated source spectra for $\gamma$-rays from dark matter, which for the annihilation channels used in this paper are similar to the one obtained in the PPPC4DMID.}
These two annihilation channels are typically chosen as representative because their spectra enclose most of the shapes of the other annihilation channels \citep{Cirelli2011}.
Electroweak corrections are taken into account only for the $\tau^+\tau^-$ channel, as they have a minimal effect in the $b\Bar{b}$ channel \citep{Ciafaloni+2011}.
The `$J$-factor' ($J$) is a geometrical quantity obtained by integrating the squared density ($\rho_\chi^2$) of DM along the line of sight (l.o.s., $\ell$) and the solid angle ($\Delta\Omega$)
\begin{equation}
    J = \int_{\Delta\Omega} \int_{\rm l.o.s.} \rho_\chi^2 \, d\ell\, d\Omega .
\end{equation}
While the DM density profile of a system can be inferred from a dynamical Jeans analysis of its member stars \citep[e.g.,][]{Bonnivard+2015, PaceStrigari2019}, spectroscopic measurements are currently unavailable for many faint systems, making a direct determination of the $J$-factor impossible.
Several previous \gray studies \citep[e.g.,][]{Drlica-Wagner+2015, Albert+2017, mcdaniel2023} have estimated the $J$-factors for targets that lacked direct measurements through scaling relations with the kinematic or photometric properties of the system \citep[e.g.,][]{Drlica-Wagner+2015, Evans:2016, PaceStrigari2019}.
In this analysis, we use the most recent versions of these relations from \citet{PaceStrigari2019}, 
\small
\begin{equation}\label{eq:photo_j}
    \frac{J_{\mathrm{photo}} (\degs{0.5})}{\mathrm{GeV}^2 \, \mathrm{cm}^{-5}} 
    \simeq 10^{18.17}
    \left( \frac{L_V}{10^4 L_\odot} \right)^{0.23}
    \left( \frac{d}{100\,\mathrm{kpc}} \right)^{-2}
    \left( \frac{r_{1/2}}{100 \mathrm{pc}} \right)^{-1}
\end{equation}
\normalsize
to estimate the $J$-factor from photometric properties.
In this equation, $d$ is the distance to the system, $r_{1/2}$ is the azimuthally-averaged physical half-light radius, and $L_V$ is the $V$-band luminosity.
Since none of the targets in our sample have confidently measured velocity dispersions, the $J$-factors used in this analysis are computed from Eq.~\ref{eq:photo_j} with the exception of UMa\,III, for which we use the value obtained by \citet{Crnorgorvcevic+2023} from the velocity dispersion scaling relation from \citet{PaceStrigari2019}, rather than the photometric one in Eq.~\ref{eq:photo_j}. 
\begin{figure*}
    \centering
    \includegraphics[width = 0.9\textwidth]{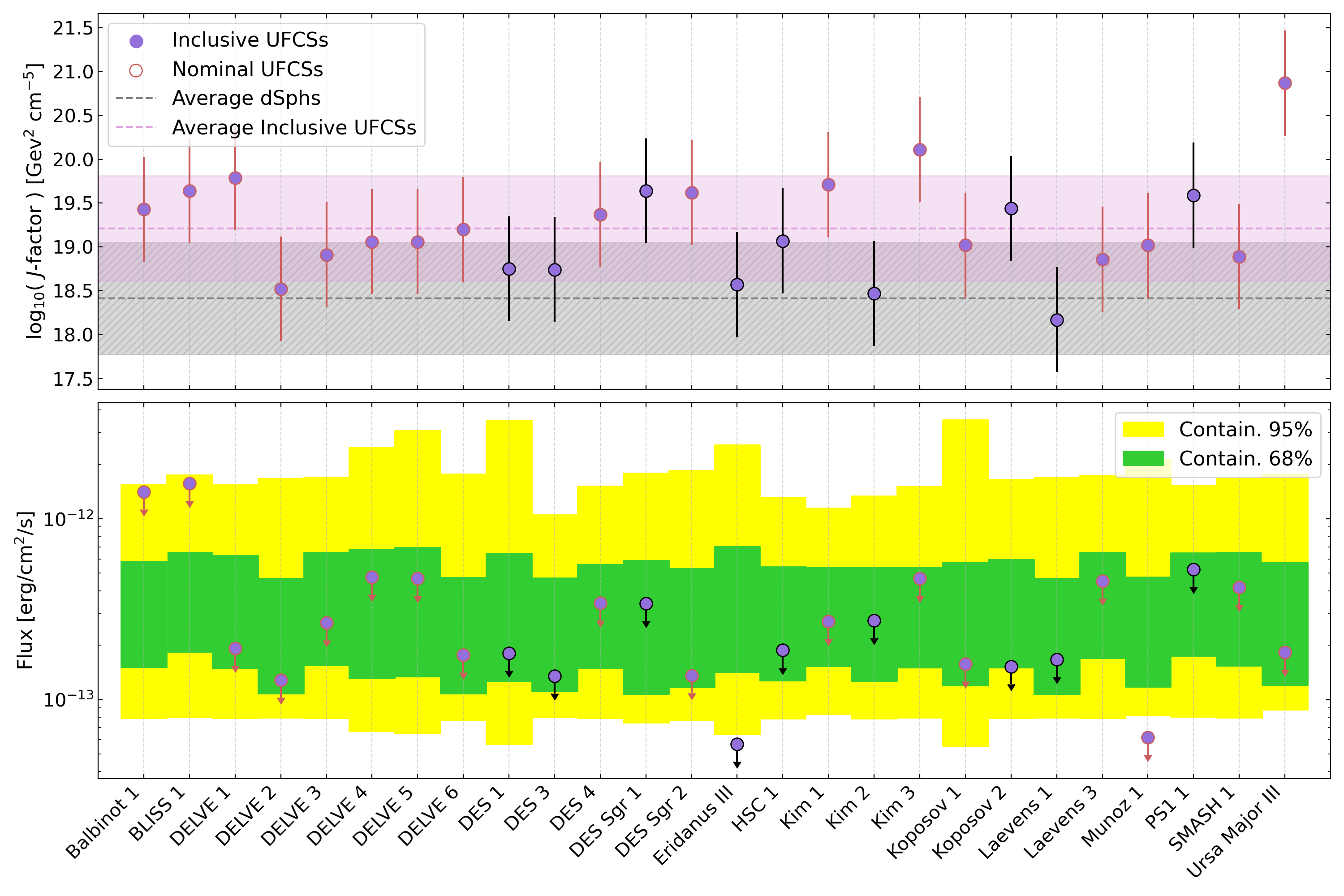}
    \caption{\textbf{Top:} $J$-factor for the selected UFCSs derived using the scaling relations of \citet{PaceStrigari2019}. We use the photometric $J$-factor relation (Eq.~\ref{eq:photo_j}) for all the sources except UMa\,III, which is instead derived from the velocity dispersion scaling relation of \citet{PaceStrigari2019} as reported by \citet{Crnorgorvcevic+2023}. We included the average $J$-factors, with their uncertainties, for the dSphs (grey band) and the inclusive sample of UFCSs (purple band) \textbf{Bottom:} \gray flux upper limits in the [0.5 GeV - 1 TeV] energy band at 95\% confidence level for the selected UFCSs. The green and yellow containment bands are obtained by selecting blank fields that reside at the same Galactic latitude as each target (within $\pm$ 5\textdegree).}
    \label{fig:J-factor_UL-flux}
\end{figure*}

In previous dSph analyses \citep[e.g.,][]{Albert+2017, mcdaniel2023}, the uncertainty on the $J$-factors estimated this way is assumed to be 0.6 dex, to represent the expected measurement uncertainty after kinematic observations.
For the UFCSs, this choice may be less conservative, since the faintness of some of these systems will likely increase the statistical uncertainty of kinematic measurements.
However, \cite{Albert+2017} found that changing the assumed $J$-factor uncertainty had a small effect compared to the uncertainty arising from the unknown nature of the UFCSs (i.e., some may be devoid of DM). \cite{Crnorgorvcevic+2023} compare the uncertainty on the sensitivity obtained from UMa III using different uncertainties for the $J$-factor (namely, the value of $\sim$1.5 dex proposed by \citealt{Errani2023}, and 0.6 dex, typically used for the dSphs -- see their Fig. 1). This has a minor effect on the sensitivity profile and its uncertainty compared to the fact that some of the UFCSs may be devoid of DM entirely, which we discuss in more detail in Sec.~\ref{sec:ResDisc}.
Consequently, we adopt the value for the $J$-factor uncertainty of 0.6 dex assumed in \cite{Albert+2017} and \cite{mcdaniel2023}, to reflect our hypothesis of similarity to the dSphs, also assuming that future velocity dispersion measurements made on the UFCSs will reach a precision similar to that of the dSphs.
In Fig.~\ref{fig:J-factor_UL-flux} we show the assumed $J$-factors and assumed uncertainties for each target in our sample\footnote{These values are based on the assumption that the UFCSs are similar in nature to the dSphs. If the subhaloes are stripped down to their stellar components, their J-factors could be lower by about an order of magnitude}.

\section{\fermi-LAT Data Analysis}\label{sec:analysis}
We analyze data from the \fermi-Large Area Telescope \citep[LAT;][]{Atwood:2009, Ajello:2021} taken between August 4, 2008 and December 1, 2022 (14.3 years). 
Our data set and analysis pipeline closely follows the dSph analysis performed by \citet{mcdaniel2023}, which makes use of the \texttt{Fermitools} (v2.2.0) via the \texttt{fermipy} (v1.2) interface \citep{Wood+2017}.
Photons from the P8R3\_SOURCE\_V3 class are selected with energies between 500 MeV and 1 TeV.
Events observed at an angle from the zenith of the spacecraft greater than 100\textdegree\, are removed to avoid contamination from Earth's limb.
Data from all four point spread function (PSF) event classes are selected and used in a joint-likelihood analysis.
This approach splits the photon events into PSF classes and includes additional information about the event-wise quality of the angular reconstruction and uses dedicated instrument response functions for each PSF event class \citep{Ackermann+2015}.

The first step of the analysis is to define a $10^{\circ} \times 10^{\circ}$ region of interest (ROI) centred on each UFCS target. 
Events are divided into eight logarithmically spaced bins of energy and spatial bins of 0.08\textdegree. 
The ROI is modeled including the Galactic diffuse emission (\texttt{gll\_iem\_v07.fits}), the isotropic spectrum for the PSF type that is being considered (\texttt{iso\_P8R3\_SOURCE\_V3\_PSF\{}\textit{i}\texttt{\}\_v1.txt}, with \textit{i} going from 0 to 3) and point-like and extended sources from the 4FGL-DR3 \citep[\texttt{gll\_psc\_v29.fits},][]{Abdollahi+2020, Abdollahi+2022}.
In particular, all sources that are up to 15\textdegree\, from the target are included to account for \gray emission originating outside the ROI.
Each UFCS target is modeled as a point-like source with a power-law spectrum.
These assumptions were made to allow for a close comparison to the dSph results from \citet{mcdaniel2023}.
Previous studies on the effects of the spatial extension of targets on DM limits have been performed for the dSphs, and have shown that modeling the sources as extended can lead to weaker limits in the DM parameter space \citep{DiMauro:2022hue}.
However, this additional uncertainty is subdominant to the uncertainty coming from our lack of knowledge of the DM density profile of each UFCS in the sample.
The model is optimized keeping as free parameters the photon index and normalization of the Galactic diffuse emission and the normalization of the isotropic component, as well as the normalization of all sources with test statistics\footnote{The TS of the sources is defined as ${\rm TS} = 2\log(\mathcal{L}/\mathcal{L}_0)$, where $\mathcal{L}$ is the likelihood derived including the target of interest in the model fit, and $\mathcal{L}_0$ is the likelihood of the null hypothesis (i.e., fixing the flux of the target source at zero).} TS $\geq 25$ that are up to 5\textdegree\, away from the target, and the normalization and photon index of all sources with TS $\geq$ 500 that are up to 7\textdegree\, away from the target.
The \texttt{find\_sources()} method is used to look for additional sources in the ROI and if any is found with ${\rm TS} > 16$, it is included in the model.
The new sources closest to a target, both with TS $\sim 20$, are found with an offset of $\degs{0.51}$ and $\degs{0.22}$ from DELVE 2 and DELVE 3, respectively\footnote{Including or removing DELVE 3 from the analysis has negligible effects on the combined constraints.}. Neither source overlaps with the target within their 95\% localization countour ($\sim \degs{0.1}$).
All the other new sources are found at an offset $>\degs{1}$.
The next step in the analysis is to calculate the spectral energy distribution (SED) for each target through the use of the \texttt{fermipy sed()} function \citep[for more information on this method, see][]{Ackermann:2014}.
Fits are performed independently in each energy bin, with the target modeled by a power-law spectrum with a fixed index of 2 and free normalization, while leaving the diffuse background normalizations free to vary.
This procedure yields a likelihood profile, $\mathcal{L}(d\Phi_{\gamma}/dE, E)$, in flux--energy space. 
The likelihood for a given DM mass and cross-section can be then computed by replacing $d\Phi_{\gamma}/dE$ with the theoretical \gray yield from DM annihilation (Eq.~\ref{eq:DM_flux}) -- which is a function of DM mass, cross-section and energy -- then summing over energy bins.
This can be summarized by the following expression:
\begin{equation}
    \mathcal{L}(\sv, M_\chi) = \sum_{E_i} \mathcal{L}\left[ \frac{d\Phi_\chi}{dE}(\sv, M_\chi, E_i), E_i \right],
\end{equation}
This likelihood profile is used to define a TS profile for the target,
\begin{equation}
    {\rm TS}(\sv, M_\chi) = 2 \left[\frac{\mathcal{L}(\sv, M_\chi)}{\mathcal{L}_0}\right],
\end{equation}
where $\mathcal{L}_0$ is the likelihood for the null hypothesis (i.e., no \gray source).
The parameter space considered covers a mass range of $M_\chi \in [5; 10^4]$ GeV and a cross-section range of $\sv \in [10^{-28}; 10^{-22}]\; \mathrm{cm}^3\mathrm{s}^{-1}$, which is motivated by GeV--TeV scale thermal relic WIMP DM models and the constraining capability of \fermi-LAT observations.
To incorporate uncertainties in the $J$-factors the \fermi-LAT likelihood function is multiplied by a $J$-factor likelihood function, $\mathcal{L}_{J}$:
\begin{align}
    \nonumber \mathcal{L}_J(J) =& \frac{1}{\ln(10) \sqrt{2\pi\sigma_J}J_{obs}}\\
                      &\times \exp{\left[-\left(\frac{\log_{10}(J) - \log_{10}(J_{obs})}{\sqrt{2}\sigma_J}\right)^2\right]}.
\end{align}
The $J$-factor likelihood function is defined as a Gaussian in the $\log J$ space, where $J_{obs}$ is the $J$-factor value estimated from the scaling relations in \citet{PaceStrigari2019}, and $\sigma_J$ is the uncertainty on the $J$-factor, assumed to be 0.6 dex in this analysis.

The limits obtained from the \gray data coincident with UFCSs can be compared to statistical expectations of the background using a \textit{`blank-field'} analysis \citep[e.g.,][]{Ackermann:2014}.
The analysis of blank fields accounts for the effects of undetected sources in the \fermi-LAT data and for the uncertainty in the models of the diffuse background emission by sampling of regions of the sky that contain no known \gray sources or likely \gray emitters based on spatial coincidence with the \fermi-LAT and multiwavelength catalogs.
The blank fields are randomly selected at high Galactic latitude ($|b| > \degs{15}$) by applying similar criteria to those used to select the sample of UFCSs---i.e., excluding regions centered within the 95\% confidence radius of a 4FGL-DR3 source or within $\degs{0.1}$ from any source in the BZCat, CRATES, and WIBRaLS catalogs.
We use the same 1000 regions selected for the dSphs analysis in \citet{mcdaniel2023}.\footnote{The data are taken from the public figshare page: https://figshare.com/articles/dataset/24058650}
From these regions, sets of 26 blank fields (i.e., the same size as the UFCS sample) are randomly selected without replacement $10^4$ times to perform a combined blank-field analysis.
We refer to \citet{mcdaniel2023} for more details on the blank-field analysis.

\section{Results and Discussion} \label{sec:ResDisc}
We analyzed ${\sim} 15$ years of \gray from \fermi-LAT data coincident with a selection of UFCSs that are potentially DM dominated. 
If their nature is confirmed, the UFCSs are expected to put stringent constraints on DM properties, since their relative proximity could lead to a higher flux of $\gamma$-rays from DM annihilation compared to the previously studied dSphs.
However, no significant \gray emission is observed in the combined sample of UFCSs.
We present the upper limits for the \gray flux from each source in Fig.~\ref{fig:J-factor_UL-flux}, with the respective values reported in Tab.~\ref{tab:sample}.

\begin{figure*}[t!]
   \centering
    \includegraphics[scale = 0.44]{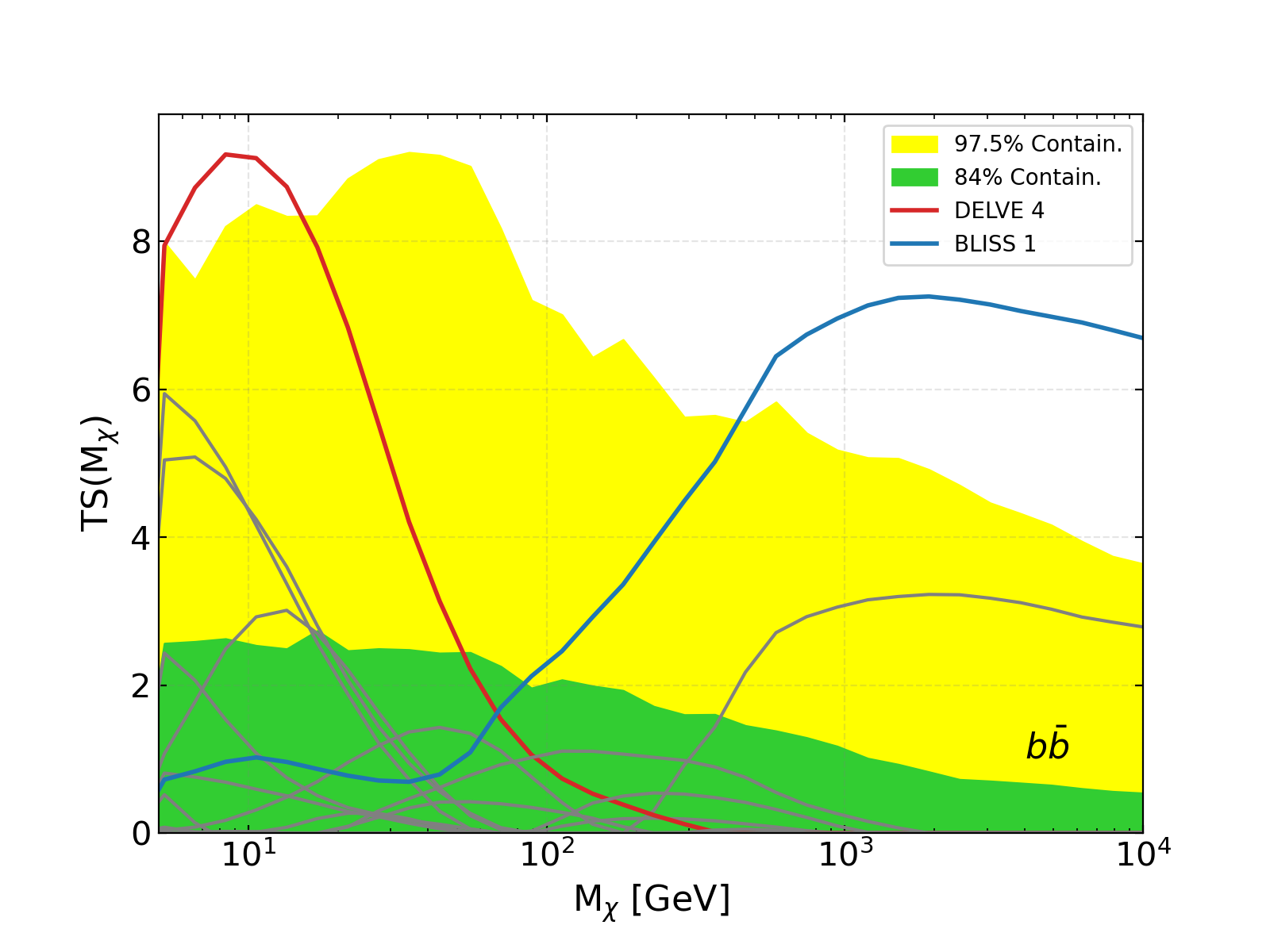}
    \includegraphics[scale = 0.44]{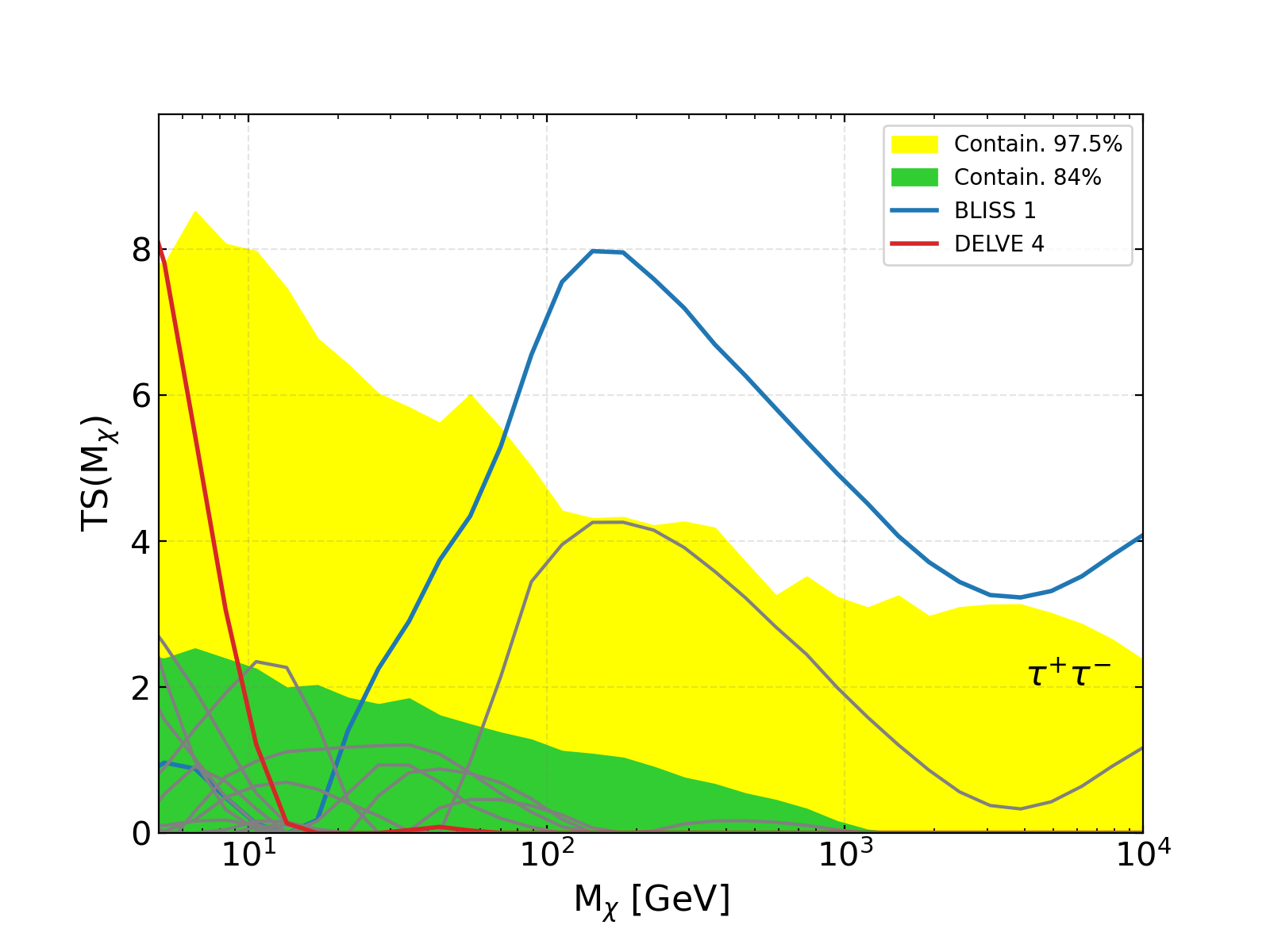}
    \caption{Maximum TS over all cross-section values vs. Mass for the individual UFCSs in the $b\Bar{b}$ (\textbf{left}) and $\tau^{+}\tau^{-}$ (\textbf{right}) channels. The colored lines highlight the targets that show a TS excess over the 97.5\% and 84\% containment regions for the individual blank fields (green and yellow bands).}
    \label{fig:TS_vs_M}
\end{figure*}

To assess the significance of a signal from the individual targets in our sample, we evaluate the maximum TS over all cross-section values as a function of the mass and compare it to the 84\% and 97.5\% containment bands from the 1000 individual blank fields (see Fig.~\ref{fig:TS_vs_M}).
Only DELVE 4 and BLISS 1 (colored lines in Fig.~\ref{fig:TS_vs_M}) show local significance $\gtrsim 2\sigma$ over the background, though we notice that these peaks in significance occur in different ranges of $M_\chi$, which would not be expected if they could both be attributed purely to DM annihilation.

\begin{figure*}[h!]
   \centering
        \includegraphics[scale = 0.42]{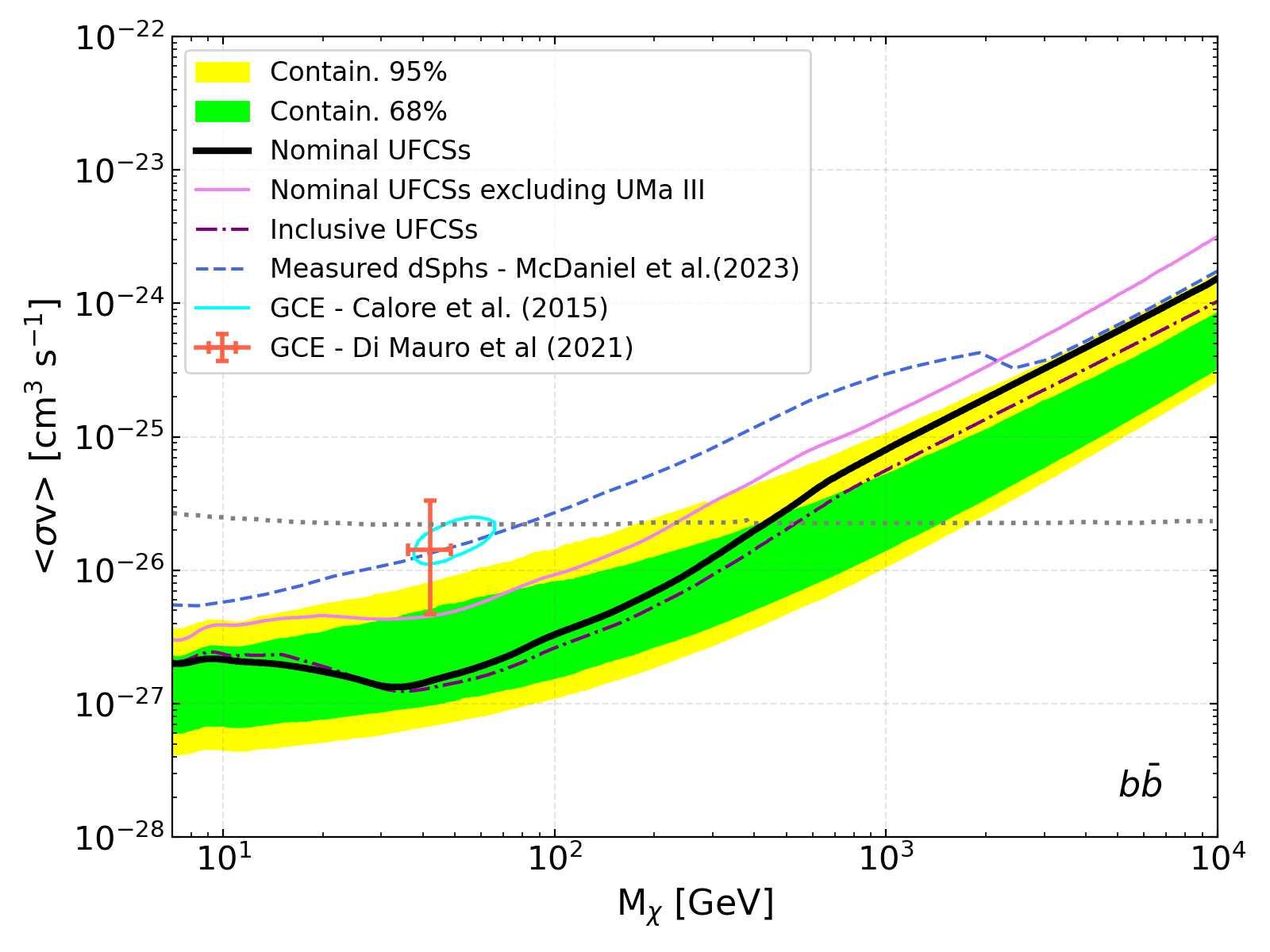}
        \includegraphics[scale = 0.42]{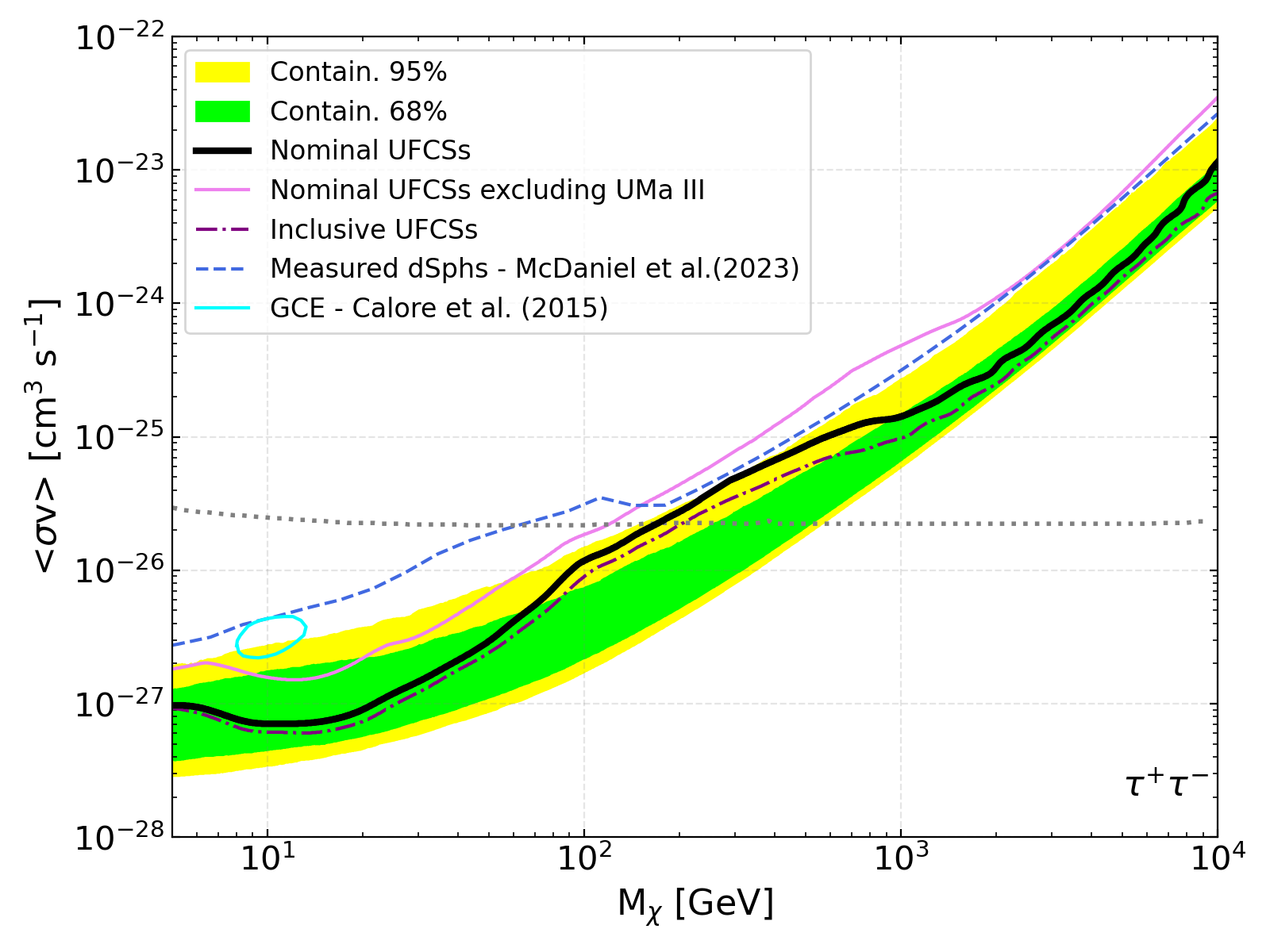}
        \includegraphics[scale = 0.42]{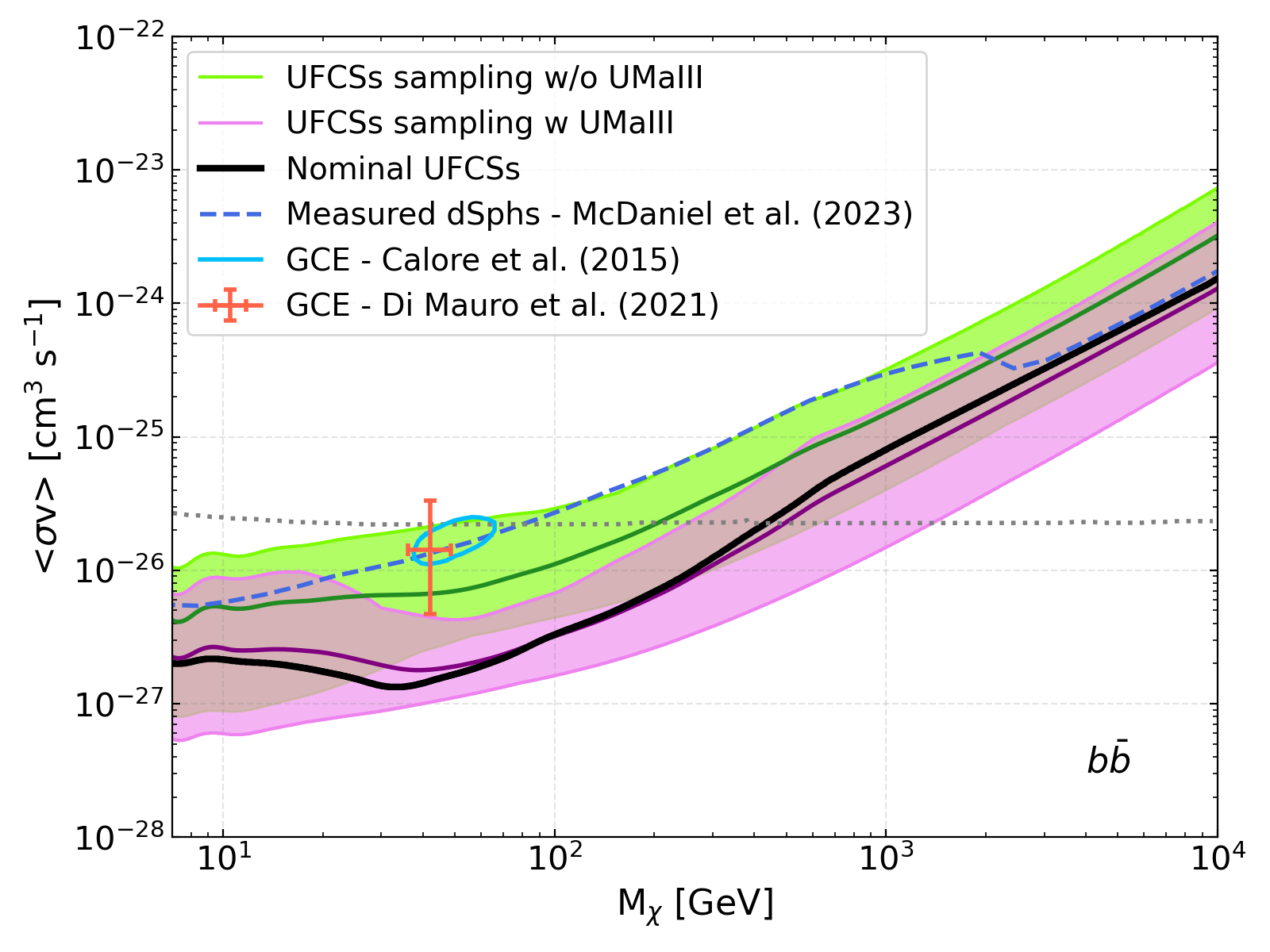}
        \includegraphics[scale = 0.42]{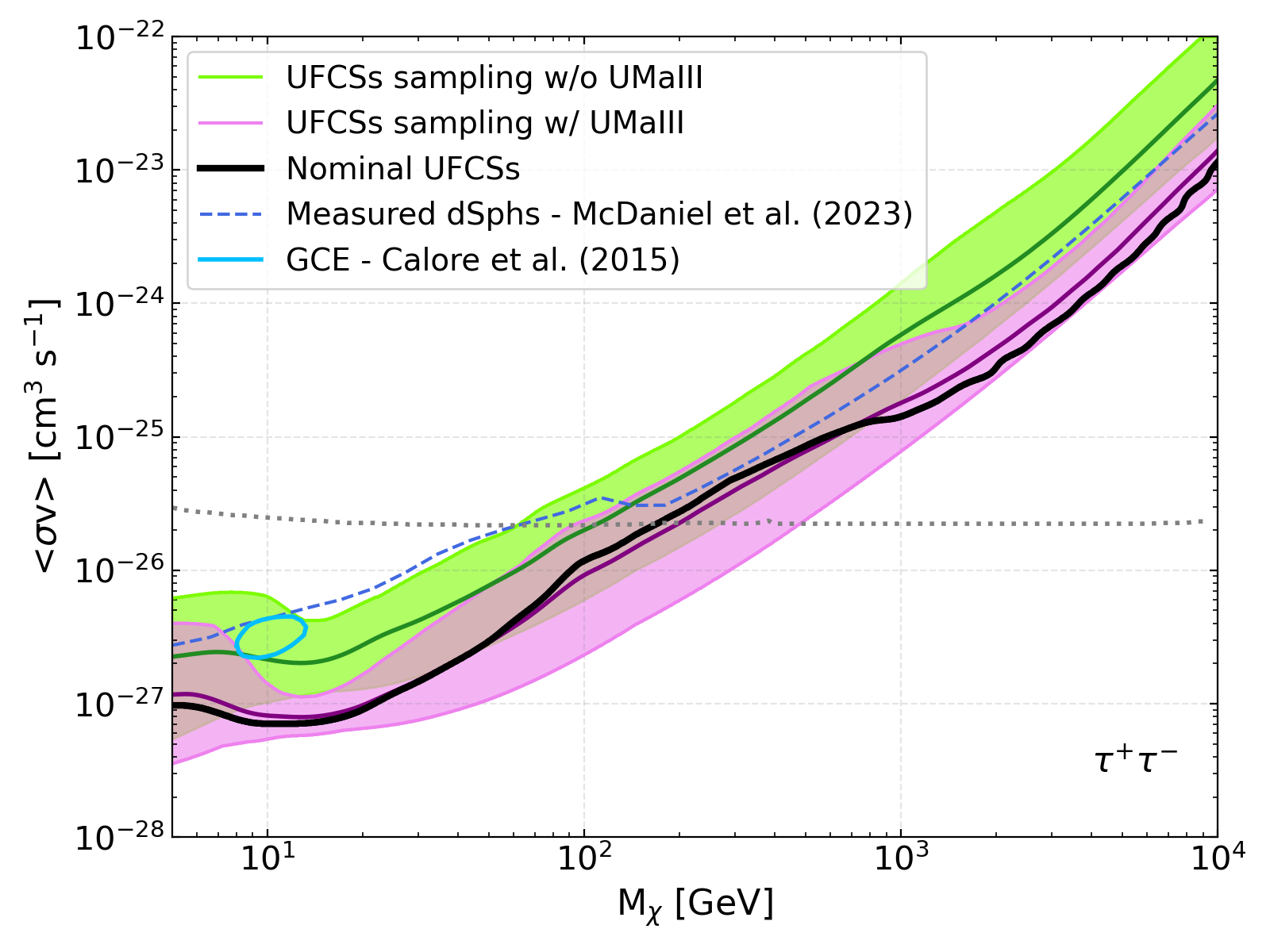}
        \includegraphics[scale = 0.42]{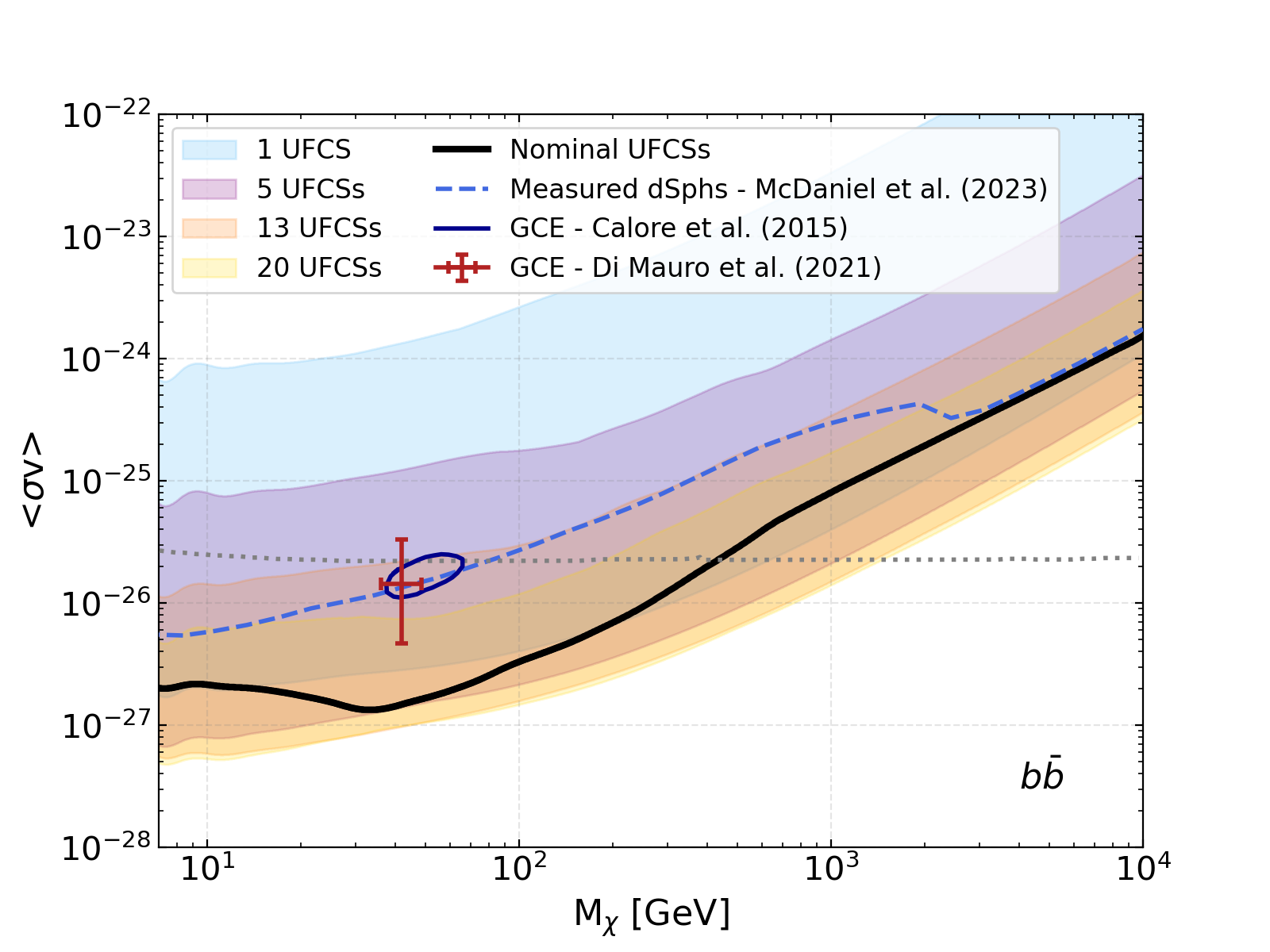}
        \includegraphics[scale = 0.42]{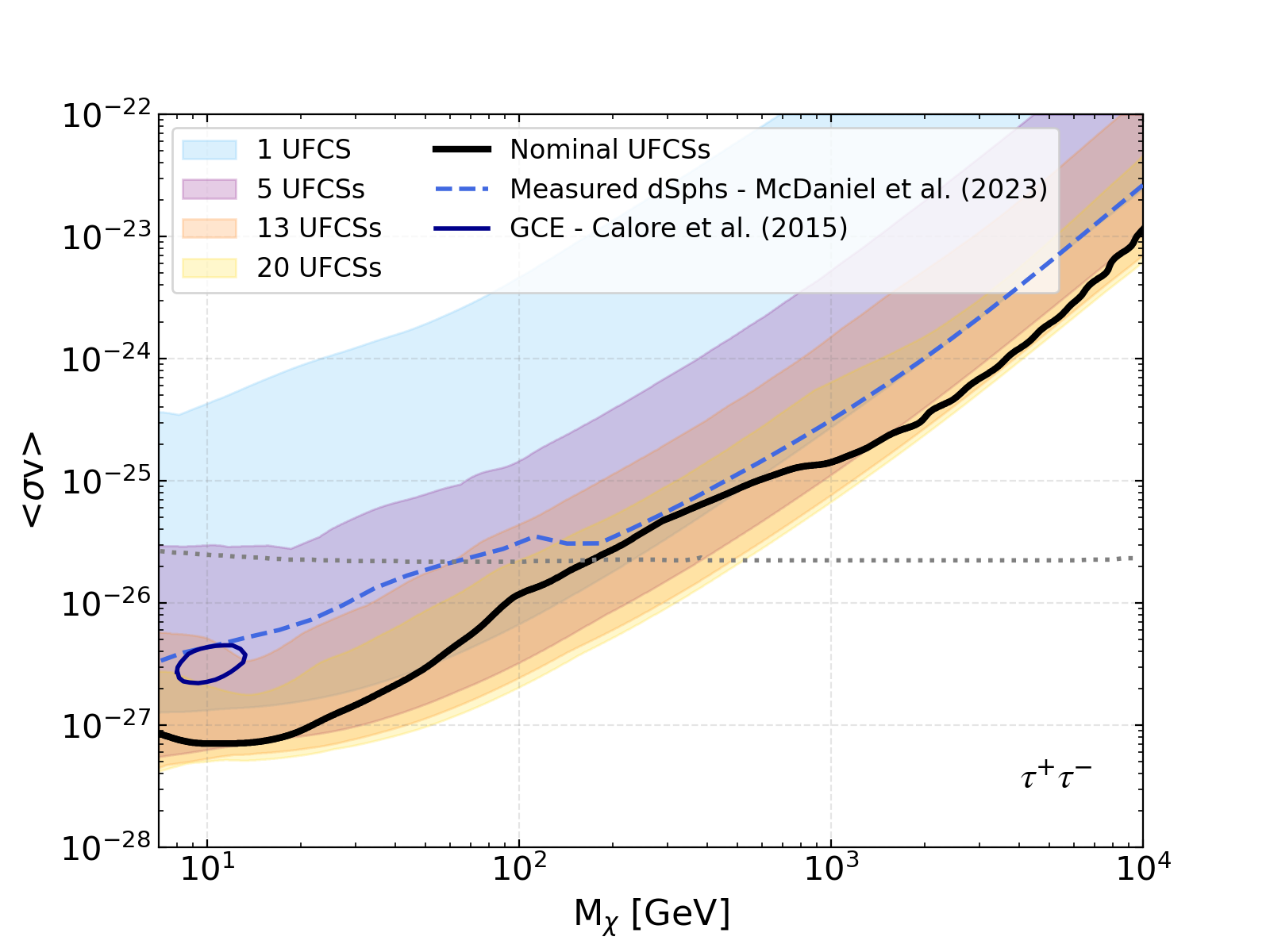}
\caption{\footnotesize \textbf{Top:} Sensitivity of \fermi-LAT observations of UFCSs to DM annihilation via the $b\Bar{b}$ (\textbf{left}) and $\tau^{+}\tau^{-}$ (\textbf{right}) channels. The solid black lines represent the constraints obtained from the combined analysis of the UFCSs in the nominal sample. The pink lines are constraints obtained excluding UMa\,III from this selection. The dot-dashed purple lines are the constraints obtained from the combined analysis of UFCSs in the inclusive sample. The yellow and green regions are, respectively, the 95\% and 68\% containment bands for the inclusive sample obtained from the combined analysis of random subsets of 26 of the 1000 blank-fields. The dotted line is the thermal relic cross-section from \citet{Steigman+2012}. The dashed blue lines are constraints from \citet{mcdaniel2023} for the measured sample of dSphs. The light blue profile represents the DM interpretation of the GCE from \citet{Calore+2015}, while the red point with error bars is the GCE measurement from \citet{DiMauro+2021_2}\\ \textbf{Center:} Sensitivity of \fermi-LAT observations to DM annihilation via the $b\Bar{b}$ (\textbf{left}) and $\tau^{+}\tau^{-}$ (\textbf{right}) channels derived by randomly selecting half of the UFCSs in the inclusive sample. The green band represents the selections that do not include UMa\,III, while the pink band represents the selections that include this source. The solid lines represent the average constraint from the respective band of the same color. The solid black line represents the constraints from the nominal sample of UFCSs. We also include results from previous analyses on the dSphs \citep{mcdaniel2023} and GCE \citep{Calore+2015, DiMauro+2021_2}.\\
\textbf{Bottom}: Sensitivity of \fermi-LAT observations to DM annihilation via the $b\Bar{b}$ (\textbf{left}) and $\tau^{+}\tau^{-}$ (\textbf{right}) channels derived for different sizes of the random subsamples.}
\label{fig:UL_profiles}
\end{figure*}

We derive the sensitivity to the null detection of \gray emission in UFCSs to put constraints on DM properties in the velocity-averaged cross-section vs.\ mass space, for the $b\Bar{b}$ and
$\tau^+\tau^-$ annihilation channels. 
In Fig.~\ref{fig:UL_profiles} we compare the upper limits obtained from the UFCSs in the inclusive and nominal samples for both channels to the results for dSphs \citep{mcdaniel2023}, for the  Galactic center excess \citep[GCE,][]{Calore+2015, DiMauro+2021}, and to the cross-section for thermal relic DM \citep{Steigman+2012}.
The sensitivities obtained from the two selections are similar, with the nominal sample yielding a slightly less stringent constraint.
We also present the results obtained excluding UMa\,III from the nominal sample, since this source dominates the sample due to its proximity and high $J$-factor.
\cite{Crnorgorvcevic+2023} have demonstrated that, according to the current estimation of its $J$-factor, UMa III alone can put constraints on DM properties that are competitive with the most recent results from dSphs.
In this context, evaluating the effects of excluding this target from the sample is crucial for two reasons.
First, it gives an upper limit on how the results presented here would be affected if further investigation of UMa III leads to a lower estimation of its $J$-factor.
Secondly, it evaluates the contribution to the constraints from the remaining targets of the nominal sample, showing that they provide a pronounced improvement to the constraints relative to the dSphs.
The upper limits obtained from both the inclusive and nominal samples are better than the constraints obtained from previous analyses at all masses and in both channels.
Even with the exclusion of UMa III, the limits obtained from the nominal sample are lower than the dSphs constraints for the majority of the mass range.
We also take into account the effects of the background through the combined blank-field analysis, from which we derive the 68\% and 95\% containment bands in the top panels of Fig~\ref{fig:UL_profiles}.
The UFCSs limits are mostly contained within these bands, and the only slight excess is observed at the highest masses in the $b\Bar{b}$ channel when UMa\,III is excluded from the analysis (Fig~\ref{fig:UL_profiles}).

As said, our analysis is premised on the assumption that the UFCSs are DM-dominated and behave similarly to the dSphs. 
Yet, given the lack of direct spectroscopic measurements, the DM content of these systems cannot be confirmed at present, and it is possible that some of these systems are DM-deficient star clusters, or that they are `micro-galaxies'.

To account for the possibility that not all of the systems are hosted in DM subhaloes, or that some of the subhaloes are not as resilient to stripping as the dSphs, we also compute the sensitivity for random subsets of the UFCSs in the inclusive sample, effectively treating the excluded systems as devoid of DM.
This allows us to gauge the main source of uncertainty in the results, which comes from the undetermined nature of the targets in our sample.

In the central panels of Fig.~\ref{fig:UL_profiles}, we compute the constraints for 1000 subsets of 13 UFCSs selected randomly from the 26 UFCSs in the inclusive sample.
This allows us to evaluate the variability of the results due to the exclusion of some of the sources. 
We highlight the selections that contain UMa\,III as these tend to lead, on average, to better constraints compared to the selections that do not contain this system. 
Again, we compare these bands to results for the dSphs and the GCE, as well as the thermal relic cross-section.
In the $b\Bar{b}$ channel the selections that do not include UMa\,III are similar to the results from the dSphs, with their average being more constraining up to high values of the DM mass ($M_\chi \gtrsim 1$ TeV).
In the $\tau^+\tau^-$ channel, this inversion happens at about an order of magnitude lower mass ($M_\chi \sim 100$ GeV). 
Yet, once again, the constraints are compatible when considering the variability coming from the different selections.
In both channels, most of the DM limits from selections that include UMa\,III are still better than the dSphs constraints, and their average is comparable to the limits obtained from the nominal sample.
A similar comparison can be made with observations of the GCE, which lie close to the average constraint of the selections that include UMa\,III.
The extension of the uncertainty bands is related to the size of the selections, as shown in the bottom panels of Fig.~\ref{fig:UL_profiles}.
Here, we display the same evaluation described above for varying subsample sizes.\footnote{This evaluation is assuming that some of the UFCSs behave like dSphs, and the rest are DM-devoid. The uncertainties would be larger than those assumed here if the UFCSs are not hosted by large DM subhaloes à la dSphs}

This study highlights the need for further investigations of the nature of UFCSs, since confirming that even a few of the observed systems are hosted in DM haloes that have not undergone significant tidal stripping could significantly increase the sensitivity of studies on DM annihilation.
Furthermore, optical imaging surveys such as DELVE \citep{Drlica-Wagner:2021},
 UNIONS \citep{Ibata:2017}, and Rubin LSST \citep{Ivezic:2019} are likely to discover more UFCSs \citep[e.g.,][]{Hargis:2014, Nadler:2020, Manwadkar+2022}.
On-going spectroscopic observing campaigns with the Keck and Magellan telescopes, as well as with the Dark Energy Spectroscopic Instrument \citep[DESI,][]{DESI+2016}, should be able to provide initial kinematic measurements following procedures similar to those described in \citet[][]{Simon+2019}; however, comprehensive measurements of the DM content of very faint systems will likely require 30-meter class telescopes \citep[e.g.,][]{GMT:2018}.

\section*{Acknowlegements}
AC, AM, CK, and MA acknowledge support from NASA grant 80NSSC22K1580 (\textit{Fermi} Guest Investigator Program Cycle 15 No.\ 151048).
Clemson University is acknowledged for generous allotment of compute time on Palmetto cluster.
ADW acknowledges support from NASA contract NNG17PZ02I (\textit{Fermi} Guest Investigator Program Cycle 9 No.~91201) and NSF Grant No.\ AST-2307126.
M.D.M. acknowledges support from
the Research grant TAsP (Theoretical Astroparticle Physics) funded by Istituto Nazionale di Fisica Nucleare (INFN). 
MASC was supported by the grants PID2021-125331NB-I00 and CEX2020-001007-S, funded by MCIN/AEI/10.13039/501100011033, by “ERDF A way of making Europe”, and the MULTIDARK Project RED2022-134411-T.

The \textit{Fermi} LAT Collaboration acknowledges generous ongoing support
from a number of agencies and institutes that have supported both the
development and the operation of the LAT as well as scientific data analysis.
These include the National Aeronautics and Space Administration and the
Department of Energy in the United States, the Commissariat \`a l'Energie Atomique
and the Centre National de la Recherche Scientifique / Institut National de Physique
Nucl\'eaire et de Physique des Particules in France, the Agenzia Spaziale Italiana
and the Istituto Nazionale di Fisica Nucleare in Italy, the Ministry of Education,
Culture, Sports, Science and Technology (MEXT), High Energy Accelerator Research
Organization (KEK) and Japan Aerospace Exploration Agency (JAXA) in Japan, and
the K.~A.~Wallenberg Foundation, the Swedish Research Council and the
Swedish National Space Board in Sweden.

Additional support for science analysis during the operations phase is gratefully
acknowledged from the Istituto Nazionale di Astrofisica in Italy and the Centre
National d'\'Etudes Spatiales in France. This work performed in part under DOE
Contract DE-AC02-76SF00515.
\software{\texttt{numpy}, \texttt{scipy}, \texttt{astropy}, \texttt{Fermitools}, \texttt{fermiPy} \citep{Wood+2017}, \texttt{dmsky},\footnote{\url{https://github.com/fermiPy/dmsky}} \texttt{local\_volume\_database},\footnote{\url{https://github.com/apace7/local_volume_database}}}

\clearpage
\begin{deluxetable*}{ccccccccc}
\label{tab:sample}
\tablecaption{List of the UFCSs that fall in the selection regions highlighted in \ref{fig:mv_hlr}. The nominal sample includes the sources that fall below the surface brightness line $\mu = 25 \mathrm{\;mag\;arcsec}^{-2}$. The inclusive sample contains all the sources in the nominal sample, with the addition of the sources in the region corresponding to $24 \mathrm{\;mag\;arcsec}^{-2} < \mu \leq 25 \mathrm{\;mag\;arcsec}^{-2}$. The values of the J-factor listed here are computed using the photometric scaling relation in Eq. \ref{eq:photo_j} \citep{PaceStrigari2019} for all systems except UMa\,III, where we adopt the value evaluated by \cite{Crnorgorvcevic+2023}. 
The J-factors and the upper limits for \gray flux at 95\% confidence level in the [0.5 GeV - 1 TeV] energy range for each source are reported in Fig.~\ref{fig:J-factor_UL-flux}.
While we report the main reference for each source in the last column, refer to the Local Volume Database \citep{Pace:2024} for a more complete list.}
\tablehead{
\colhead{Name} & \colhead{GLon} & \colhead{GLat} & \colhead{Distance} & \colhead{$R_{1/2}$} & 
\colhead{$M_V$} & \colhead{$\log_{10}(J_{photo})$} & \colhead{$F_{UL}$} & \colhead{Ref.}\\
\colhead{} & \colhead{[\textdegree]} & \colhead{[\textdegree]} & \colhead{[kpc]} & \colhead{[pc]} &
\colhead{} & \colhead{[$\mathrm{GeV}^{2}/\mathrm{cm^5}$]} & \colhead{ [$\times10^{-12}$ erg/$\mathrm{cm}^{2}$/s]} & \colhead{}
}
\startdata
        \hline
           \textbf{Nominal sample}\\
        \hline
        Balbinot 1 & 75.18 & -32.64 & 31.9 & 5.57 & -1.2 & 19.43 & 14.07 & \citet{Balbinot+2013}\\
        BLISS 1 & 290.83 & 19.65 & 23.7 & 4.14 & 0.0 & 19.64 & 15.68 & \citet{Mau+2019}\\
        DELVE 1 & 14.19 & 30.29 & 19.0 & 6.08 & -0.2 & 19.79 & 1.92 & \citet{Mau+2020} \\
        DELVE 2 & 294.24 & -47.79 & 71.0 & 21.48 & -2.1 & 18.52 & 1.28 & \citet{Cerny+2023} \\
        DELVE 3 & 335.85 & -27.06 & 56.0 & 6.52 & -1.3 & 18.91 & 2.66 & \citet{Cerny+2023} \\
        DELVE 4 & 42.31 & 56.43 & 45.0 & 6.41 & -0.2 & 19.06 & 4.76 & \citet{Cerny+2023} \\
        DELVE 5 & 19.38 & 61.36 & 39.0 & 7.71 & 0.4 & 19.06 & 4.68 & \citet{Cerny+2023} \\
        DELVE 6 & 290.57 & -49.08 & 79.8 & 0.43 & -1.2 & 15.61 & 1.76 & \citet{Cerny+2023b} \\
        DES 4 & 270.87 & -33.44 & 31.3 & 7.56 & -1.1 & 19.37 & 3.43 & \citet{Torrealba+2019} \\
        DES Sgr 2 & 163.58 & -52.20 & 23.8 & 11.04 & -1.1 & 19.62 & 1.36 & \citet{Luque+2017} \\
        Kim 1 & 68.52 & -38.42 & 19.8 & 6.91 & 0.3 & 19.71 & 2.71 & \citet{Kim+2015} \\
        Kim 3 & 310.86 & 31.79 & 15.1 & 2.29 & 0.7 & 20.11 & 4.69 & \citet{Kim+2016} \\
        Koposov 1 & 260.97 & 70.76 & 48.3 & 8.71 & -1.0 & 19.02 & 1.58 & \citet{Munoz+2018} \\
        Laevens 3 & 63.60 & -21.18 & 61.4 & 11.43 & -2.8 & 18.86 & 4.53 & \citet{Longeard+2019}\\
        Munoz 1 & 105.44 & 45.58 & 45.0 & 6.41 & -0.4 & 19.02 & 0.62 & \citet{Munoz+2018} \\
        SMASH 1 & 292.14 & -27.99 & 57.0 & 9.45 & -1.0 & 18.89 & 4.18 & \citet{Martin+2016}\\
        Ursa Major III & 194.61 & 73.68 & 10.0 & 3.0 & 2.2 & 20.87 & 1.83 & \citet{smith2023}\\
        \hline
           \textbf{Inclusive sample}\\
        \hline
        DES 1 & 310.52 & -67.83 & 76.0 & 5.42 & -1.4 & 18.75 & 1.81 & \citet{Conn+2018} \\
        DES 3 & 343.83 & -46.51 & 76.2 & 6.21 & -2.0 & 18.74 & 1.35 & \citet{Luque+2018} \\
        DES Sgr 1 & 142.83 & -75.79 & 26.5 & 2.71 & 0.3 & 19.64 & 3.40 & \citet{Luque+2017} \\
        Eridanus III & 274.95 & -59.60 & 91.0 & 8.34 & -2.0 & 18.57 & 0.57 & \citet{Conn+2018} \\
        HSC 1 & 66.32 & -41.84 & 46.0 & 5.89 & -0.2 & 19.07 & 1.87 & \citet{Homma+2019}\\
        Kim 2 & 347.15 & -42.07 & 100.0 & 13.96 & -3.3 & 18.47 & 2.74 & \citet{Kim+2015b} \\
        Koposov 2 & 195.11 & 25.55 & 34.7 & 4.44 & -0.9 & 19.44 & 1.52 & \citet{Munoz+2018} \\
        Laevens 1 & 274.81 & 47.85 & 145.0 & 21.51 & -4.8 & 18.17 & 1.67  & \citet{Laevens+2014}\\
        PS1 1 & 10.04 & -17.42 & 29.6 & 4.74 & -1.9 & 19.59 & 5.25 & \citet{Torrealba+2019} \\
\enddata
\end{deluxetable*}
\newpage
\bibliographystyle{aasjournal}
\bibliography{biblio}
\end{document}